\documentclass[preprint,12pt]{elsarticle}

\usepackage{amssymb}

\journal{Chemical Physics Letters}

\def\hipco{HiPco}
\def\swnt{single-walled carbon nanotubes}

\def\cnt{carbon nanotubes}
\def\nt{nanotubes}

\def\vhs{Van Hove singularities}
\def\cm-1{cm$^{-1}$}

\def\sph{sp$^3$}
\def\celsius{$^{\circ}$C}
\def\comocat{CoMoCat}
\def\gic{graphite intercalation compounds}

\def\graf{graphite}
\def\thf{tetrahydrofuran}

\def\tgms{thermogravimetry-mass spectrometry}
\def\c60{C$_{60}$}

\begin{document}

\begin{frontmatter}

\title{Breakdown of diameter selectivity in a reductive hydrogenation reaction of \swnt}

\author[Wigner Research Centre for Physics]{Katalin Nemeth}
\address[Wigner Research Centre for Physics]{Institute for Solid State Physics and Optics, Wigner Research Centre for Physics, Hungarian Academy of Sciences, P.O. Box 49, H 1525 Budapest, Hungary}
\author[Institute of Materials and Environmental Chemistry]{Emma Jakab}
\address[Institute of Materials and Environmental Chemistry]{Institute of Materials and Environmental Chemistry,
Research Centre for Natural Sciences, Hungarian Academy of Sciences, P.O. Box 286, H 1519 Budapest, Hungary}
\author[Wigner Research Centre for Physics]{Ferenc Borondics\fnref{cls}}
\fntext[cls]{Present address: Soleil Synchrotron,
BP 48, L'Orme des Merisiers, 91192 Gif sur Yvette C\'{e}dex, France}
\author[Wigner Research Centre for Physics]{Hajnalka M. T\'oh\'ati}
\author[Wigner Research Centre for Physics]{\'Aron Pekker\fnref{ucr}}
\fntext[ucr]{Present address: Center for Nanoscale Science and Engineering, Departments of Chemistry and Chemical \& Environmental Engineering, University of California, Riverside, CA 92521, U.S.A.}
\author[Wigner Research Centre for Physics]{M\'onika Bokor}
\author[Wigner Research Centre for Physics]{Tam\'as Vereb\'elyi}
\author[Wigner Research Centre for Physics]{K\'alm\'an Tompa}
\author[Wigner Research Centre for Physics,Obuda University]{S\'andor Pekker}
\address[Obuda University]{\'Obuda University, Doberd\'o \'ut 6, H 1034 Budapest, Hungary}
\author[Wigner Research Centre for Physics]{Katalin Kamar\'as\corref{cor1}}
\cortext[cor1]{Corresponding author}
\ead{kamaras.katalin@wigner.mta.hu}

\begin{abstract}
Reductive hydrogenation was applied to two types of  \swnt\ (SWNTs) with different diameter range. Alkali metal intercalation, followed by reaction with methanol, led to hydrogenated products. Both yield and selectivity of this reaction showed strong dependence on diameter, contrary to expectation based on simple curvature effects. The observed yield, as detected by \tgms\ and $^1$H-NMR, is drastically reduced in small-diameter tubes where the alkali dopant does not reach the inside of the bundles. Wide range optical transmission measurements were employed to determine the selectivity and indicate that besides higher yield, lower diameter selectivity occurs above a critical diameter.
\end{abstract}


\end{frontmatter}


\section{Introduction}
\label{}

Chemical modification of carbon nanotube surfaces by sidewall
reactions is important for any application where further chemical bonding or increase in solubility
is desired. Nanotube chemistry, however, constitutes many challenges compared to conventional organic chemistry methods: pristine \nt\ are insoluble in most solvents, and solid samples contain bundles of many tubes, held together by van der Waals forces. To increase contact between reactants and the tube surface, exfoliation and solubilization are required. These unique characteristics also mean that other factors than the chemical properties have to be taken into account when predicting the yield of a chemical reaction.

Reductive reactions, where the first step consists of charging the reactants, are especially well suited for exfoliation. Several types of modified Birch reduction are found in the literature \cite{Chen98,Pekker01,Borondics05,Borondics07},
using ammonia, \thf\ and ethylenediamine as solvent. The first step is electron transfer onto the \nt, where repulsion between the resulting negatively charged "supermolecules" helps exfoliation and promotes further reactions. Similar processes can happen in alkali metal intercalated graphite compounds \cite{Dresselhaus81}, resulting in partially hydrogenated graphite \cite{Bergbreiter78}.

Following \gic \cite{Dresselhaus81}, nanotube exfoliation possibilities and processes \cite{Dailly05,Tanaike09}, we applied direct reduction by intercalating alkali metals into carbon nanotube bundles. Based on these results and the rather similar reactivity of \graf\ and \cnt,
we expected alkali metal intercalation to exfoliate and reduce
nanotube bundles. By this method, the step of carbanion formation during the reductive modification could be separated in space and time from the attachment of the functional group (unlike other Birch-type reactions) \cite{Voiry10}.

Adding alkali metals to \cnt\ in excess, over time a saturation concentration (about KC$_9$ for potassium) is reached \cite{Iwasa01}. A stable phase (with composition KC$_{27}$ determined by XPS) is formed in a few minutes at 180 \celsius. Structural characterization of K- and Cs- doped SWNTs by X-ray diffraction and electron microscopy proved that the alkali ions are attached to the tube surfaces, increasing the bundle size, instead of being encapsulated into the tubes \cite{Bower98}. Sidewall functionalization of the nanotubes can proceed from the doped phase by electrophilic addition.

The general selectivity of sidewall reactions in nanotubes is believed to be determined principally by structural strain caused by $\pi$-orbital misalignment on the curved surface \cite{Niyogi02}, resulting in higher reactivity  of smaller diameter tubes. Considering one nanotube, these features determine diameter selectivity thermodynamically.
Taking into account, however, that realistic nanotube samples consist of bundles, there are other important phenomena
that must be considered, such as kinetics, steric effects and
energetics of all processes and intermediate products in a
reaction.

The effect of bundling is an unusual and exciting feature of nanotube properties and reactions, and has been subject of many previous investigations. One especially relevant to reductive reactions is the careful and detailed study by Kukovecz et al. \cite{Kukovecz03} on the first step of these reactions, the doping of bundled nanotubes, using resonant Raman spectroscopy and conductivity measurements. They found that in \hipco\ tubes a minimum in doping level appears at around 1.1 nm. They explained their findings by taking into account the morphology of the bundles, and calculated the required relative lattice expansion as a function of the size of the intertube channels in the bundles, which is proportional to the tube diameter.
According to their calculations, above 1.3 nm tube diameter, there is no need for lattice expansion for incorporating K$^+$ ions. At low diameters, however, another effect has to be taken into account, the increasing stacking interaction in the bundle, inversely proportional to tube diameter \cite{Henrard99}. This latter effect competes with the increasing reactivity at higher curvature.

In this paper, we describe the results of intercalation of SWNTs by alkali metals, followed by electrophilic addition of protons to yield hydrogenated products. We used two types of commercial nanotube samples as starting materials, with different diameter range, above and below 1.2 nm. As dopant ions we used potassium and rubidium, similar in chemical properties but differing in size, in order to reach two extreme combinations (potassium with the larger diameter tubes, rubidium with the smaller ones). The products were investigated by \tgms\ (TG-MS), $^1$H-NMR spectroscopy, Raman spectroscopy and wide range optical transmission spectroscopy to obtain detailed information about the reactivity through the composition, thermal and optical properties of the samples, with special attention to diameter selectivity. We find that above a critical intertube channel/ionic radius ratio, the yield increases and becomes independent of diameter.

\section{Experimental}

Two types of commercially available \swnt, differing in average diameter, were used: P2 by Carbon Solutions and CoMoCat CG by SouthWest NanoTechnologies. In order to reach two extremes of the dopant radius to diameter ratio r$_D$/d$_{NT}$, P2 tubes were doped with potassium and CoMoCat tubes with rubidium. Parameters of the reactants are summarized in Table \ref{tbl:nanotubes}. Calculated intertube channel radii from the mean diameter values in Table \ref{tbl:nanotubes}, according to Ref. \cite{Kukovecz03}, are 0.15 nm for P2 and 0.096 nm for CoMoCat. Comparing these values with the dopant radii, the K/P2 combination represents a case where the channels can easily accomodate the dopant ions, whereas for Rb/CoMoCat the ionic radius of the dopant by far exceeds the intertube channel radius. This difference is expressed in the r$_D$/d$_{NT}$ ratio, given in the last column of Table \ref{tbl:nanotubes}.

\begin{figure}
  \includegraphics[width=10cm]{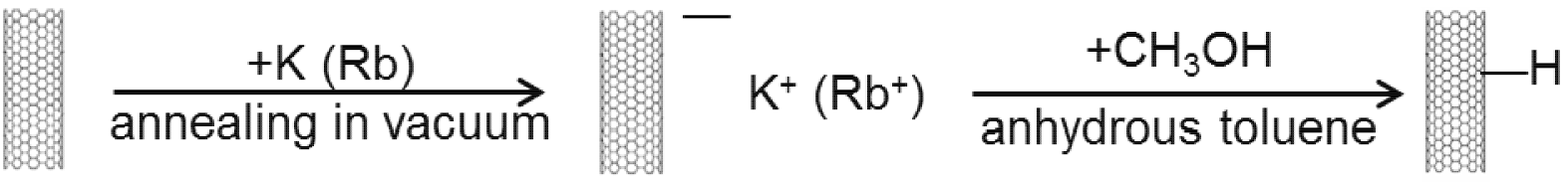}
  \caption{Direct reduction of \swnt\ by intercalating potassium}
  \label{sch:directred}
\end{figure}

The hydrogenation method, which was inspired by \gic, is
described below and shown in Figure \ref{sch:directred}.

\begin{table}
  \caption{Specific parameters of nanotube reactions.}
  \label{tbl:nanotubes}
  \begin{tabular}{|c|c|c|c|c|c|}
    \hline
    Nanotube &  Diameter & Mean & Dopant & Ionic & r$_D$/d$_{NT}$ \\
        &  range (nm) & diameter (nm)  &  & radius (nm) &\\
    \hline
    P2  & 1.2-1.7 & 1.60 & K & 0.138 & 0.086 \\
    \comocat & 0.57-1.17 & 0.90 & Rb & 0.152 & 0.169 \\
    \hline
  \end{tabular}
\end{table}

In the case of P2, about 100 mg of as-received SWNT was first annealed in dynamic vacuum (at ~10$^{-6}$ mbar) at 250 \celsius\ for 12 hours, followed by transfer into an argon dry box. Potassium was added in a glass vial, keeping the SWNT-K molar ratio 4:1. The glass vial was sealed on a vacuum line. Annealing at 200 \celsius\ for 12 hours was enough for potassium to intercalate the \nt. Intercalation was indicated by the copper/gold color of the sample. Subsequently, the sample was put into a Schlenk-type flask with a funnel. 40 ml anhydrous toluene (Sigma-Aldrich, cryo-distilled from Na-K alloy) was added to the flask and 20 ml into the funnel. Toluene was used as a totally aprotic solvent to avoid side reactions with any other H source, and to control alkali metal atoms to intercalate the \nt\ and not to dissipate in the solvent. Sonication was applied for 15 minutes to enhance the intercalation process. Next, 5 ml methanol (Sigma-Aldrich, used as received) was filled carefully into the funnel. Methanol was added dropwise into the flask during sonication. Sonication was continued for 2 more hours, and the mixture was left overnight. The sample was filtered on a \textit{Millipore} nylon membrane filter (0.1 $\mu$m pore size), washed with ethanol, 1:3 HCl:H$_2$O, distilled water, ethanol and acetone. Finally it was dried in dynamic vacuum at 200 \celsius\ for 12 hours. The product obtained this way was transferred back into the dry box. The whole process described above, except prior annealing, was repeated two more times in order to investigate whether it is possible to improve the degree of hydrogenation by applying successive steps. For \comocat, rubidium was used instead of potassium as intercalating agent, in order to achieve larger bundle expansion \cite{Liu03}. The main products of these reactions are hydrogenated \nt, but there are side reactions, such as hydrogen evolution, when attachment of H to the nanotube is kinetically hindered, or when the unreacted alkali metal reduces methanol directly.

Reference samples were made of pristine P2 and \comocat\ by performing the same steps as with the hydrogenation reactions (initial annealing, annealing in sealed glass tube, addition of methanol, washing, annealing in dynamic vacuum), \textit{except addition of alkali metal.}

Thermogravimetry-mass spectrometry (TG-MS) measurements were used to determine the composition of the samples, particularly the H content. The main purpose of the measurement is the quantitative determination of the decomposition products as a function of temperature. Mass change with temperature is directly measured by a Perkin-Elmer TGS-2 thermobalance and a HIDEN HAL 2/301 PIC quadrupole mass spectrometer. 2-4 mg sample in a Pt vessel was heated up to 800 \celsius\ with 20 \celsius/min rate in Ar atmosphere. A portion of the volatile products was introduced into the mass spectrometer (operated at 70 eV in electron impact ionization mode) through a heated glass-lined steel capillary. Ion intensities were normalized to $^{38}$Ar isotope of the carrier gas to eliminate errors resulting from the shift in MS intensities. To measure hydrogen (\textit{m/z} 2), a calibration with TiH$_2$ is necessary. During the measurements, signals of 16 ions were followed.

Wide line $^1$H-NMR measurements and data acquisition were
accomplished by a Bruker AVANCE III NMR spectrometer at the
frequency of 82.4 MHz with a stability better than $\pm10^{-6}$. The inhomogeneity of the magnetic field was 2 ppm. Free induction decays (FIDs) were measured at room temperature. Known amounts (weight) of the nanotubes (typically 7-15 mg) and  adamantane (99+\%, Sigma-Aldrich) were put in
Teflon capsules. The FID measured on the empty capsule was
subtracted from the FID of the actual capsuled sample to correct for background. The amplitude of the FID at zero time is proportional to the number of $^1$H nuclei in the sample \cite{Tompa03}. The first 9-10 $\mu$s of the FID was lost in the dead time of the spectrometer. The observed FIDs were extrapolated back to zero time by fitting Gaussian functions to obtain its zero-time amplitude. The FID of adamantane was used for calibration in calculating the hydrogen concentrations. There were residual magnetic catalyst particles in the samples (typically 2-5 w/w\%), but they did not disturb the measurement significantly \cite{Borondics05}.

Raman spectra were taken by a Renishaw 1000B spectrometer using 785 nm excitation wavelength, with 4 cm$^{-1}$ spectral resolution. The laser power was kept sufficiently low in order to exclude heat damage.

Wide range optical transmission measurements were carried out on self-supporting thin films made of the samples \cite{Wu04}.
Transmission data between 25-52500 \cm-1\ were recorded, by a Bruker IFS 66v/s FT-IR instrument in  the far and mid-infrared region, a Bruker Tensor37 in the near infrared, and a Jasco v550 spectrometer in the visible and UV. Optical conductivity data were obtained by performing Kramers-Kronig (KK) transformation on the transmission data (atomic force microscopy was used to determine sample thickness for KK calculations) \cite{Pekker06}. Optical conductivity spectra were fitted by the Drude-Lorentz model. After subtracting the background and all other peaks related to transitions between different \vhs, we obtained the contribution of the specific peaks \cite{Pekker11}.

\section{Results and discussion} \label{sec:results}

\begin{figure}
  \includegraphics[width=7cm]{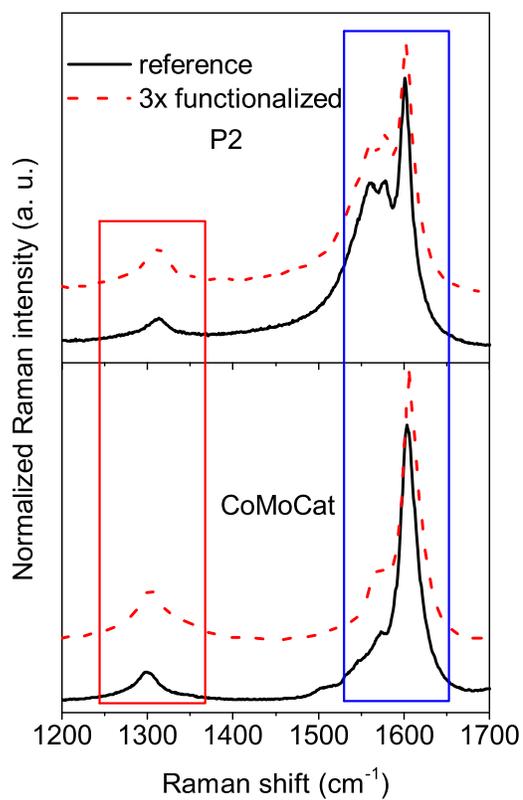}
  \caption{Raman spectra of the two samples (excitation wavelength: 785
  nm) before functionalization (reference) and after three functionalization steps. Red frame: D modes; blue frame: G modes. Each spectrum is normalized to the G mode. Increasing $D/G$ ratio indicates sidewall hydrogenation.}
  \label{fgr:ramandg}
\end{figure}

Raman spectra were recorded in order to detect changes in the $D/G$ mode intensity ratio. The increase in $D/G$ ratio indicates the presence of more defects in the nanotube sidewall. This is characteristic of functionalized \nt, in which the \sph\ carbon atoms in the sidewalls act as defects \cite{Graupner06,Mueller10}. Raman spectra of both samples are shown in Figure \ref{fgr:ramandg}. Each spectrum is normalized to the $G$ band to be comparable. The $D/G$ ratio increases upon functionalization, but the change cannot be used for quantitative estimation of the sidegroup content, because of the complicated origin of the D band \cite{Reich04}. To determine the hydrogen content more precisely, TG-MS and NMR spectrometry were employed.

\begin{figure}
  \includegraphics[width=7cm]{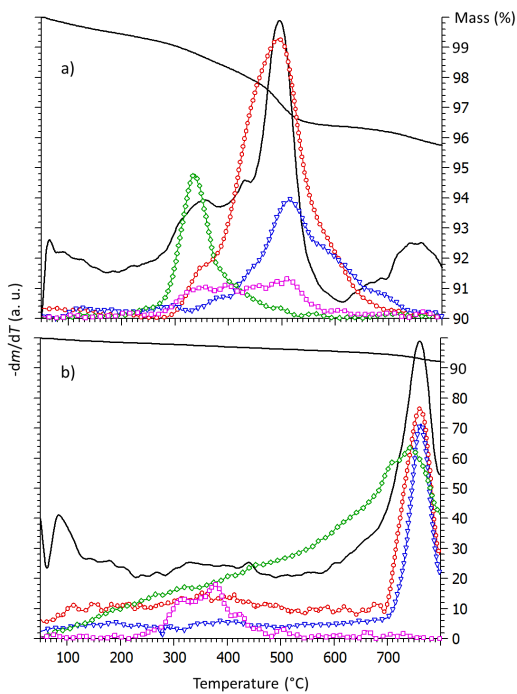}
  \caption{Typical TG-MS curves of the samples. a) P2, b) \comocat. Black solid lines in both graphs represent mass loss and its derivative, respectively. Notations for graph a): (red circles) m/z 2 hydrogen; (blue triangles) m/z 16 methane; (green diamonds) m/z 31 methanol; (magenta squares) m/z 92 toluene. For graph b): (red circles) m/z 2 hydrogen; (blue triangles) m/z 28 CO; (green diamonds) m/z 44 CO$_2$; (magenta squares) m/z 92 toluene, where m is the atomic mass and z the atomic number.}
  \label{fgr:tgms}
\end{figure}

Results of the TG-MS measurements are shown in Figure \ref{fgr:tgms} and Table \ref{tbl:Hdata}. In case of the \comocat\ tubes the H-content was so small that the hydrogen evolution peak faded into the background, therefore the exact temperature of hydrogen evolution could not be determined.

\begin{table}
  \caption{H content and thermal stability of the samples. NMR data are related to the samples with the highest degree of hydrogenation measured by TG-MS. H content determined by NMR are corrected with the H content of the reference samples to correct with the H content of other H containing materials (toluene etc.).}
  \label{tbl:Hdata}
  \begin{tabular}{ccccccc}
    \hline
    Sample&$T_{\textrm{max}}$ [H] (\celsius)&\multicolumn{3}{c}{H /100C (TG-MS)}&[H] (mmol/g) &H /100C \\
     & & & & & (NMR) & (NMR) \\
    \hline
    &&\multicolumn{3}{c}{Sample number}&&\\
    &&1&2&3&&\\
    \hline
    P2&350&2.17&2.41&3.61&4.63$\pm$0.27&5.6\\
    \comocat&--&$<$1&$<$1&$<$1&2.39$\pm$0.15&2.9\\
    \hline
  \end{tabular}
\end{table}

The degrees of hydrogenation measured by NMR follow the same trend as TG-MS: the value obtained for \comocat\ is smaller than that for P2. Both methods proved the presence of hydrogen, although they have not yet been systematically compared. Since NMR measures the total hydrogen content in the sample, we consider the TG-MS method as yielding more precise quantitative results of hydrogen with chemical bonding to the nanotube walls. The hydrogen content of the products formed from P2 is between 2-4 at\%. In contrast, in smaller diameter \comocat\ samples, where Rb was used as intercalating alkali metal, the degree of hydrogenation is very small. This fact is in good agreement with previous studies \cite{Bendiab08,Harley04}.

From special features in the optical spectra, conclusions can be drawn about the reactivity of the starting material \cite{Mueller10,Pekker08,Nemeth10}. The energy of the interband transitions (or likewise, transitions between excitonic levels close to the band edge) scales with the inverse diameter of the respective tubes. The width of the peaks in the optical spectra reflects mainly the diameter distribution of peaks in the sample, and therefore the change in shape upon a chemical reaction reflects the change in distribution between the starting material and the product. The most precise procedure to extract these changes from solid-state spectra is to compare the optical conductivity. This quantity is additive when several independent processes are involved (like light absorption by different nanotubes) and its calculation takes into account the reflectance at the interfaces, which can heavily influence the optical density calculated from transmittance \cite{Pekker06}. The model we used \cite{Pekker11} involves a reliable background correction and results in curves reflecting one type of transition across various diameter samples. The advantage of using optical spectra to determine changes in diameter distribution over Raman spectroscopy is that there is no resonance process which prefers certain nanotubes over others with selective increase in the scattering intensity.

\begin{figure}
  \includegraphics[width=7cm]{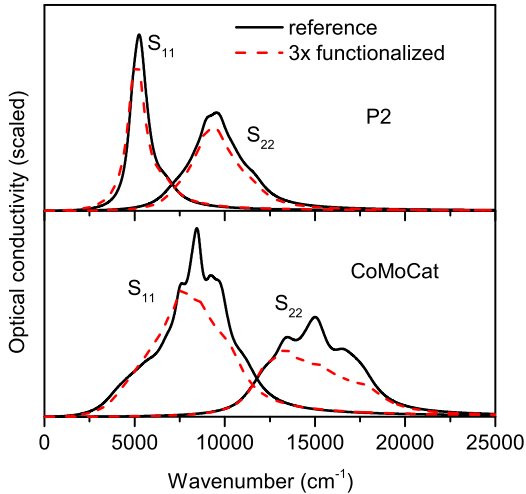}
  \caption{The effect of hydrogenation on the first (S$_{11}$) and second (S$_{22}$) semiconducting interband transitions in the optical spectra. In both cases, the reference sample and the product with the highest hydrogen content are shown.}
  \label{fgr:opt}
\end{figure}

Figure \ref{fgr:opt} compares the change in optical conductivity upon hydrogenation in the spectral regions connected to the first (S$_{11}$) and second (S$_{22}$)  transitions between semiconducting nanotube energy levels. The curves shown are those of the products with the highest hydrogen content, obtained after three steps of hydrogenation. Intensity loss due to hydrogenation is independent of frequency in P2, and increases with increasing frequency in \comocat. This change indicates that the reactivity is determined by the penetration of the dopant into the bundles: in P2, the intensity decreases independently of frequency, whereas in \comocat\ the inverse diameter dependence returns. In the latter case, the effect of the bundles seems to disappear, since the alkali atoms will cover only the bundle surface. This effect results in much lower yield but "normal" diameter selectivity, determined by the curvature of the nanotubes.

In the first approximation, intercalation should proceed according to the first transition energy (S$_{11}$ in semiconducting nanotubes). Since this quantity scales inversely with diameter, the first reaction step would be more probable for higher diameter tubes. These two opposite trends (that of the ionic intercalation step and the chemical reaction, described above) determine the rate of the reaction and therefore it is not surprising that it is not monotonous with diameter. An added complication arises because the intercalation happens inside bundles, introducing additional factors. Since intercalation
into bundles formed by larger diameter tubes is less hindered, these bundles will be more reactive, both by their larger electron density and better availability for reactant molecules \cite{Kukovecz03}. Above the diameter limit estimated  (1.3-1.5 nm), where the intertube channels are wider than the ionic radius, potassium intercalation becomes unhindered and because of high availability for reactant molecules (loosened bundles) very little diameter selectivity can be detected. This is the
case of the K/P2 sample. As the channels become narrower compared to the intercalant ion, intercalation becomes more and more hindered until it will be energetically totally unfavorable. In this case (Rb/\comocat), electron transfer is possible only at the surface of the bundles, which causes a drastic lowering in availability of tubes \cite{Bendiab08}. Below the transitory diameter range the diameter selectivity will be determined by the chemical reaction, favoring higher curvature, as observed in the case of \comocat.

We did not address the question of selectivity between metallic and semiconducting tubes; as in the first intercalation step all the tubes acquire metallic character due to doping with charge carriers from the alkali metal \cite{pichler99}, we do not consider this effect significant.

\section{Conclusion} \label{sec:conc}

We have synthesized hydrogenated \swnt\ by alkali metal
intercalation and subsequent reaction with methanol. By using two samples with widely different diameter distributions (P2 and \comocat), we could investigate diameter selectivity. We have revealed the role of alkali metal intercalation in diameter selectivity of similar reductive reactions. The results are in agreement with previous studies on alkali metal intercalated \nt. We demonstrated that factors other than structural strain must be considered in reactions that are driven in not totally homogeneous media such as bundled nanotubes. For practical purposes of nanotube chemistry, our results explain why a nanotube sample with low average diameter can in fact produce lower yield in chemical reactions than nanotubes above a critical diameter limit.

\section{Acknowledgement}

This research was funded by the Hungarian National Science Fund (OTKA) under grant No. 105691.

\bibliographystyle{elsarticle-num}
\bibliography{H_nanotube}

\end{document}